\documentclass[conference]{IEEEtran}
\IEEEoverridecommandlockouts
\usepackage{cite}
\usepackage{amsmath,amssymb,amsfonts}
\usepackage{algorithmic}
\usepackage{graphicx}
\usepackage{textcomp}
\usepackage{xcolor}
\usepackage{paralist}
\usepackage[linesnumbered,ruled,vlined]{algorithm2e}

\SetCommentSty{mycommfont}
\def\BibTeX{{\rm B\kern-.05em{\sc i\kern-.025em b}\kern-.08em
    T\kern-.1667em\lower.7ex\hbox{E}\kern-.125emX}}
\begin{document}

\title{Dance of the DAOs: Building Data Assets  as a Use Case\\
}

\author{\IEEEauthorblockN{ Sarad Venugopalan and Heiko Aydt}
\IEEEauthorblockA{\textit{Singapore-ETH Centre} \\
sarad.venugopalan@sec.ethz.ch, aydt@arch.ethz.ch }

}

\maketitle

\begin{abstract}
	Decentralised Autonomous Organisations (DAOs) have recently piqued the interest of participants from diverse backgrounds, including business owners, engineers, individual and institutional investors. In part, the promised autonomy (less rigid
	structure and more voice) in decision making along with ease of market access, has resulted in its participants pouring in their time and economic resources. In a DAO, governance  is typically enacted via posting proposals and collectively voting on it. The winning proposals are then implemented.
	However, governance alone may be insufficient, when its participants' economic incentives are misaligned.
	Governance and tokenomics need to work in tandem to ensure business stability.
	We present a case study on an example building data asset  from the construction industry and present its tokenomics. We show its working, both as a
	caretaker and strategic DAO, to illustrate its  effects on governance and DAO stability.
	The case study serves as an example for participants to decide whether
	their DAO tokenomics are aligned with participation incentives.
	Finally, we propose the DAO tension quadrilateral to study DAO stability and build a tool to measure agreement among its participants.
\end{abstract}


\section{Introduction}
A DAO is a decentralised multi-stakeholder organisation managed via a majority agreement. Its task is to assure the continued wellbeing of the organisations' value producing system (such as goods and services), governed via cooperation for mutual benefit.
A DAO is equipped with a platform (similar to a tamper resistant public bulletin board~\cite{SuwitoTSDTUS22}) to provide its de-centralised stakeholders a mechanism to collectively make decisions. I.e., it provides a platform for participants to conveniently and securely meet, propose, deliberate and vote on issues affecting them.
The governance mechanisms in modern DAOs are built and implemented on top of agreed upon computerised rules and contracts. These rules are enforced by the consensus implemented on a blockchain. An important feature of a DAO is decentralised trust, allowing its stakeholders to partake in governance without having to trust any one stakeholder in the organisation, when such rules are incorporated into the DAO smart contracts~\cite{Szabo1997}.

DAOs provide the much needed organisational structure required to govern a (for profit or non-profit) business, in a decentralised setting. Along with traditional organisations, the DAO aims to provide good governance, by ensuring there are no disruptions to its consumers, and every supplier to the business in its value chain is continuously incentivised to play their part.
A great deal of work has been carried out in the systemisation of knowledge in blockchain governance and to shed light  on shortcomings in existing systems~\cite{governance2022}.
However, governance alone may be insufficient to prevent participants from exiting or the business from destabilising. For stability, incentives of the individual stakeholders must be aligned with that of the organisation. These incentives are essential for business cooperation, as  tokens are  for coordination~\cite{Lamberty2020,Hunhevicz2022}.
The basis for coordination are the tokens but the structure that allows participants to coordinate based on complex rules comes from the DAO.
We demonstrate the interplay of tokenomics and DAO governance  (i.e., coordination, cooperation,  and structure) with building assets as a use case.

We make the following contributions,
\begin{compactenum}

	\item [\textit{i}.)] We present two scenarios, one where DAO instability is reduced by design, and in the other, a certain degree of instability is inherent by design (see section \ref{ssec:caretakerstrategic}). Further, we use our building asset case study, as an example, to improve DAO stability (see section~\ref{ssec:improvingstabilitybydesign}).
	
	\item [\textit{ii}.)] We propose a mechanism to study tension between individuals/groups in a DAO (see section~\ref{ssec:tensionquad}), and develop a tool to measure and quantify participant agreement (see section~\ref{ssec:measuring}).
\end{compactenum}

The rest of the paper is organised as follows.
Section~\ref{sec:prelims} gives the definitions for the building blocks in our building asset case study.
Section~\ref{sec:motivation} presents the motivation for using building data assets.
Section~\ref{secsanalysis} details the system analysis and threat model.
The system architecture is presented in section~\ref{sec:sysarch}. The addition of the DAO to the building ecosystem is detailed in section~\ref{sec:dao}.
Section ~\ref{sec:discussion} discusses the interconnection between the solutions presented.
We conclude with our findings and contributions in section~\ref{sec:conclusions}.

\section{Preliminaries}
\label{sec:prelims}
Various pieces of information combined in a meaningful way, may not only lead to new (derived) information but also to insights/knowledge. For example, information about electricity consumption combined with information about the location (which floor) and orientation (facing which side of the building?) of the units within a building may lead to insights such as --- avoid high-floor, west-facing units to avoid high electricity bills.
Information does not have a value per se. Information has value when it can be used for a particular purpose. For example, it may enable the consumer of the information to gain insights and make better-informed decisions. The nature of the value of information can be manifold but generally varies with the consumer and their specific use-case. 
One consumer may attribute a higher value to some information than another. For example, insights gained from using information about electricity consumption may help a prospective tenant to choose a unit with lower operational cost.

If an asset is “a resource with economic value that an individual, corporation, or country owns or controls with the expectation that it will provide a future benefit"~\cite{Barone2022}, then a data asset is a (digital and intangible) resource that an individual, corporation, or country owns or controls with the expectation that it will provide a future benefit. In order for data to be considered an asset it needs to satisfy the following requirements.
\begin{compactenum}
	
	\item [(R1)] It must be impossible to arbitrarily duplicate a data asset - this turns data into a (unique) and identifiable object. It also introduces the element of scarcity.
	
	\item [(R2)] It must be possible for the owner(s) of a data asset to take possession and assume exclusive control (with possible exception of a regulatory authority) - this allows assets to become private property (subject to the laws of the jurisdiction the asset may be subject to) enabling its owners to accumulate, hold, delegate, rent, or sell their property.
	
	\item [(R3)] It must be possible for a data asset to be subject to a regulatory authority (if required) - this gives the data object legitimacy in the context of any relevant jurisdiction in the physical world.
	
	\item [(R4)] It must be possible to audit a data asset (e.g., in order to verify its integrity and test for any impairments).
	
	\item [(R5)] It must be possible to generate/derive and extract useful information from a digital asset (or parts thereof) that can be purchased/licensed and consumed by third parties to their benefit, thus generating a source of revenue.
	
\end{compactenum}

A Building Data Asset (BDA) is a data asset plus a domain-specific Application Program Interface (API) to interact with and manipulate the data in ways meaningful to the domain of interest (i.e., the building domain in case of BDAs). Without a domain-specific API that gives meaning to the underlying data asset, it would merely be a bucket of bits and bytes. There can be many APIs within a domain. For example, within the building domain, one may have different APIs for different use cases such as power consumption (monthly power usage in an apartment unit), construction (recyclable material used~\cite{Honic2021}), and others. The same data asset may be accessed/interacted  through different APIs, whereby each API only interacts with some parts of the underlying data asset.

A token is a standardised unit of accounts. A token represents a particular interest in some aspect of the data asset and is used to coordinate the activities of the business. For example, it may represent a share of ownership in the data asset, or right to use.
Typically, a token represents a data asset created on a public blockchain to enable its properties. When stored on a blockchain, the  asset does not contain the bulk of the associated data stored along with it, instead only a link to it. When tokens are issued on the blockchain, and correspond to the properties of the data asset, they are called tokenised assets. It may be fungible or non-fungible, depending on its application.

\section{Motivation to use Building Data Assets }
\label{sec:motivation}
\begin{figure*}
	\centering
	\includegraphics[width=1.0\linewidth]{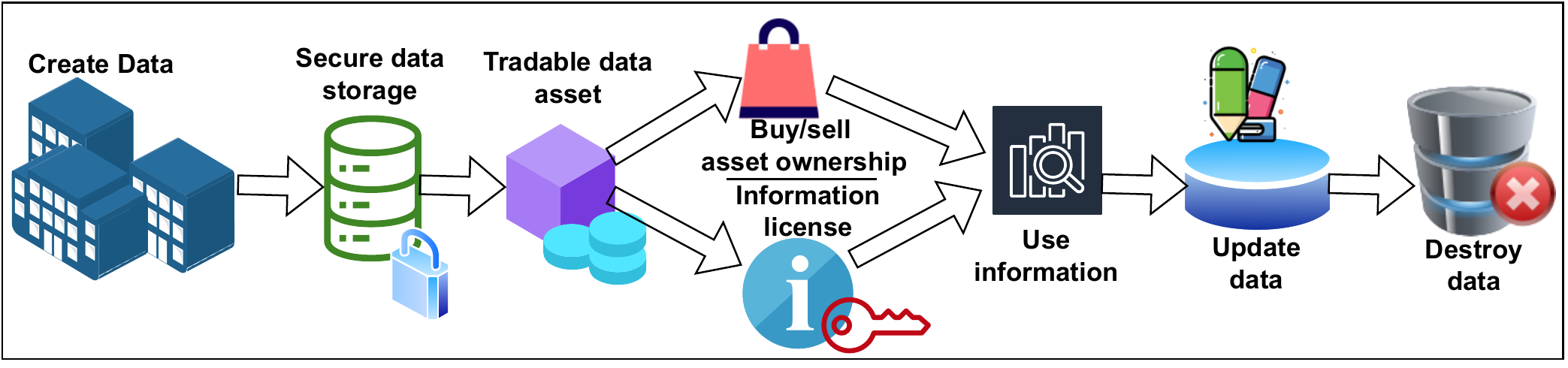}
	\caption{ Building data is first created and securely stored before it is turned into a tradable asset. The ownership of data may be traded or a data licence issued to the consumer for the use of information. The owner may use the information for profit and the consumer may use it for analysis and actionable insights. The data in storage may be updated over time, and reused. In general, the data is destroyed when the building is demolished, unless it has further value.  }
	\label{fig:lifecycle}
\end{figure*}

Building information is currently not treated as something valuable beyond its immediate scope of application. One of the reasons for this is the lack of a long term value proposition and/or clear pathways for generating revenue using building data. In order to monetise building data, it has to be first systematically generated or collected (e.g., during the construction phase of a building). Furthermore, building data must be maintained, curated and regularly updated to remain relevant and up-to-date as time passes. All this incurs costs and companies may not see the point of doing so beyond what is needed to achieve their immediate tasks at hand.

For example, design, construction and engineering companies may create and use building data (e.g., design blueprints, component databases, de-/construction details~\cite{Sanchez2021}) during the various phases of building construction. However, they may have no further use for this data after the building is constructed. However, ideally this data would be made available to businesses that specialise in operating and maintaining a building, who could also maintain and enhance the building data (e.g., with data from operations - such as energy consumption). Depending on the nature of specialisation, such businesses may also be able to monetise the data by selling insights (e.g., about maintenance needs/events) to interested third parties (e.g., building operators/owners, maintenance contractors, tenants). Yet other businesses may specialise in the mining of urban materials or by providing the necessary insights (e.g., what materials can be found where) to interested third parties - thus contributing to a circular economy~\cite{Cetin2021} of construction materials.

What if the companies (e.g., construction, design and engineering companies) that generate/collect data as by-product and that do not have the necessary infrastructure, skill (or interest for that matter) to monetise this data beyond their own use of the data (e.g., blueprints used by a construction firm during building construction) could sell/transfer their building data to interested third parties in a standardised and structured manner - just like any other asset? This may allow businesses to emerge that specialise in acquiring (and aggregating) certain building data to generate revenue by exploiting valuable insights or generating revenue by selling them. While technical standards for building data itself may already exist (e.g., Building Information Modelling), there is generally a lack of standards and methods for addressing the economic/financial aspects of data.

Pathways for generating revenue would provide the necessary incentives to maintain and update data until the building is demolished (and possibly beyond). We believe that building data assets have a crucial role to play in the context of sustainable material flows in circular future cities~\cite{Heiko2022}. Among other potential use-cases for building data assets, they may provide information about building components and their materials to entities that specialise in mining and recycling building materials, thus contributing to a circular building materials economy.

\section{System Analysis and Threat Model}\label{secsanalysis}
\subsection{System Requirements}
We define the following requirements for the Building Data Assets.

\subsubsection{Unhindered Access and Transfer of Tokenised Assets}
BDAs should not be locked into a ‘permissioned’ platform to facilitate unhindered transfer between entities, not limited to entities that are part of any particular group, consortium or otherwise.

\subsubsection{Censorship and Tamper resistance of Tokenised Assets}
BDAs should be highly censorship resistant, i.e., it should be resistant to denial  attacks (such as denial of service~\cite{Msisac2022} or disputes arising from assets entrusted to a custodian~\cite{Georgina2015}). Tamper resistance is to thwart unauthorised modifications.

\subsubsection{Atomicity of Tokenised Assets}
An asset owner may hold multiple assets related to materials and services in a building. Each asset is atomic.  For example, two  different special interest groups combined into one might not find the combined information they need under a single asset or data component.  They may have to query each of the two assets separately. This partitioning simplifies the  management of asset economics.

\subsubsection{Security of Data Assets}
Providing security is necessary for two reasons. First, a data asset is of value only when its record keeping can be secured from ownership tampering, illegal alteration and access to trade and indisputable settlement of the asset. This is essential in a public setting where information can be bought/sold. The second reason is that only data that is owned and controlled can be sold.

\subsection{Design Goals}
The following design goals are considered to meet the system requirements.
\begin{compactenum}
	
	\item Flexible asset trade and information transfer. The trade and transfer of  assets and information must be public and straightforward without compromising its security goals. A public blockchain with a decentralised exchange is used to achieve this goal. Non-fungible tokens are used to point to the location of data in a datastore.
	
	\item The data linked to the asset must be atomic. I.e., the information stored is clearly partitioned.
	
	\item Secure and censorship resistant data assets. The asset and its data must be secured against the attacks under the threat model (see section~\ref{ssec:threatmodel}) to mitigate security risks. This is achieved by using the in-built security and censorship resistance of a blockchain.

	
	\item Economic incentives to keep the wheels moving. For services to be provided and maintained, we primarily rely on sustainable incentive mechanisms for cooperation and tokens for coordination, instead of  altruism or regulatory enforcement (which might result in the very minimum done to stay compliant).
	
	\item Regulatory requirements on data. Some countries require compliance with the ‘right to forget’. I.e., people have the right to ask their personal information to be removed from the public domain. To meet this requirement, we will not store any personal information on any append only ledger or make it public.
\end{compactenum}

\subsection{Threat Model}
\label{ssec:threatmodel}

The adversary has bounded computing power
(i.e., they are unable to break used cryptographic primitives
under their standard cryptographic assumptions).
They may however collude with other Byzantine adversaries or take advantage of Byzantine faults in the blockchain network.
The adversary is assumed not to have the resources to  block communication entering the blockchain
network.
The  communication over the  channel is tamper resistant since  all transactions
sent to the blockchain are digitally signed by its sender.

The presence of Byzantine faults in our public network justifies the use of a Byzantine fault-tolerant State Machine Replication (a.k.a Blockchain).
Our datastores are secured via a suitable access control mechanism.  Only  authorised users are permitted to retrieve data.
The datastore may be made byzantine fault-tolerant by running 3.$f$+1;$f\geq 1$ instances and implementing a suitable BFT consensus. However, a study of the complexities involved when $f>1$ and potential solutions are beyond the scope of this work.

\section{System Architecture}
\label{sec:sysarch}

In this section, we present the stakeholders, components and its interconnections for our building data asset ecosystem. We also introduce the tokens used for coordination between its stakeholders.
Fig.~\ref{fig:lifecycle} shows the evolution of building data from its inception to termination. Fig.~\ref{fig:ecosystem} shows the architecture of our building ecosystem.

\subsection{Stakeholders}
\begin{figure*}
	\centering
	\includegraphics[width=1.0\linewidth]{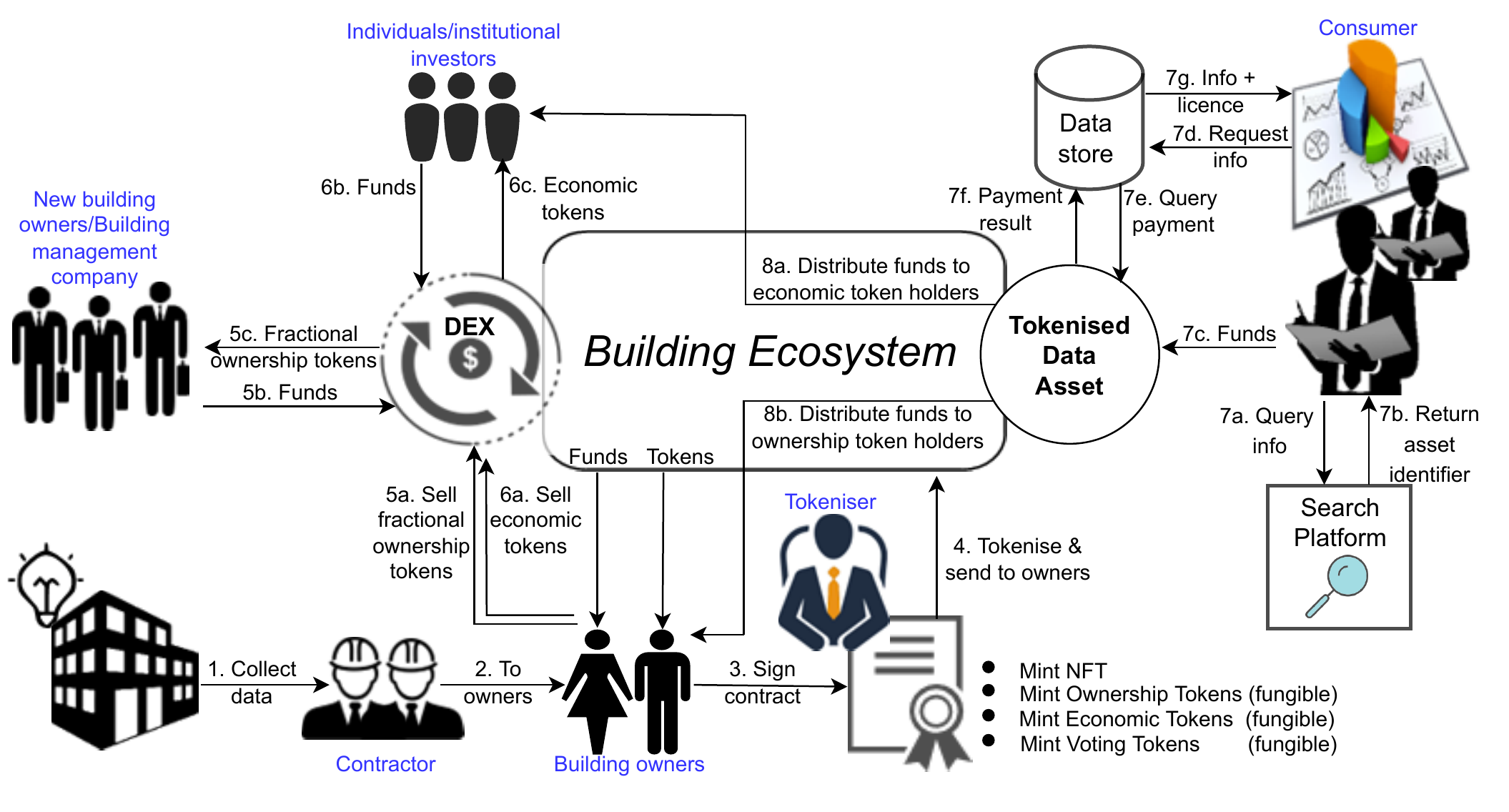}
	\caption{ The sequence of steps in bootstrapping and operating  the building ecosystem --- 1. Contractors collect building data and 2. is passed to the  building owner. 3. Building owners sign a legally binding  contract with the tokeniser. 4. Tokeniser issues/mints all the tokens agreed in the legal contract signed and sends it to the owners. Owners use a decentralised exchange (DEX) to sell tokens, 5. either to sell a part of the ownership tokens, or 6. to raise capital using economic tokens. Further, the consumers interested  in the building data are able to 7a. query on a search platform for the 7b. unique identifier of the tokenised data asset holding the required information. Once determined, 7c. the consumer pays the correct tokenised data asset. 7d. Next, the consumer requests the data store for the information. 7e. The data store queries the NFT if the required payment was made. 7f. On confirmation of payment, 7g. the requested information and licence are released to the consumer from a datastore. The payment made by the consumer to  tokenised data asset (NFT) is proportionally distributed  between 8a. economic and 8b. ownership token holders, based on the rules in the NFT smart contract.
	}
	\label{fig:ecosystem}
\end{figure*}

Using the building ecosystem steps (see Fig.~\ref{fig:ecosystem}) as a baseline, we identify the following stakeholders and service providers.
\begin{compactenum}
	
	\item Contractors are people tasked to collect the required data (e.g., building data or electricity data), digitise it in a prescribed format, and supply it to the building owner.
	
	\item Data asset owner(s) are  people who have ownership of the building data.  The data asset owners are building owners or those with close relationships with the building. Asset owners may employ contractors one-off, or periodic. Typically, data assets are owned by the building owner or bought by an information asset ownership company from a building owner.
	
	\item Data certifier(s) are one or more competent and authorised person(s) who physically verifies the correctness of building data (or its components) on behalf of the asset owners, who then  digitally signs the timestamped data (not shown in Fig.~\ref{fig:ecosystem}). 
	
	\item A tokeniser  company/entity acts as a bridge between the physical and blockchain world. They are responsible for tokenising data assets and providing those tokens to the building owner.
	
	\item Data storage providers. Two types of data storage are used --- cloud storage and public blockchain.
	Cloud storage services are hosted/rented on a subscription basis. It holds the datastores.
	The data asset is tokenised on a public blockchain.

	\item Information users/consumers  may be any individual, group, or organisation who use the data held by data asset owners via queries to profitably analyse it for useful insights. The data ownership is retained with the asset owner. The consumers are licensed to use the information but not sell it. The consumers pay for the information received.
	
	\item Investors. Any member of the public (individual or institutional) who buys tokens for dividends (royalties) and trades them.
	
\end{compactenum}

\begin{figure}
	\centering
	\includegraphics[width=0.9\linewidth]{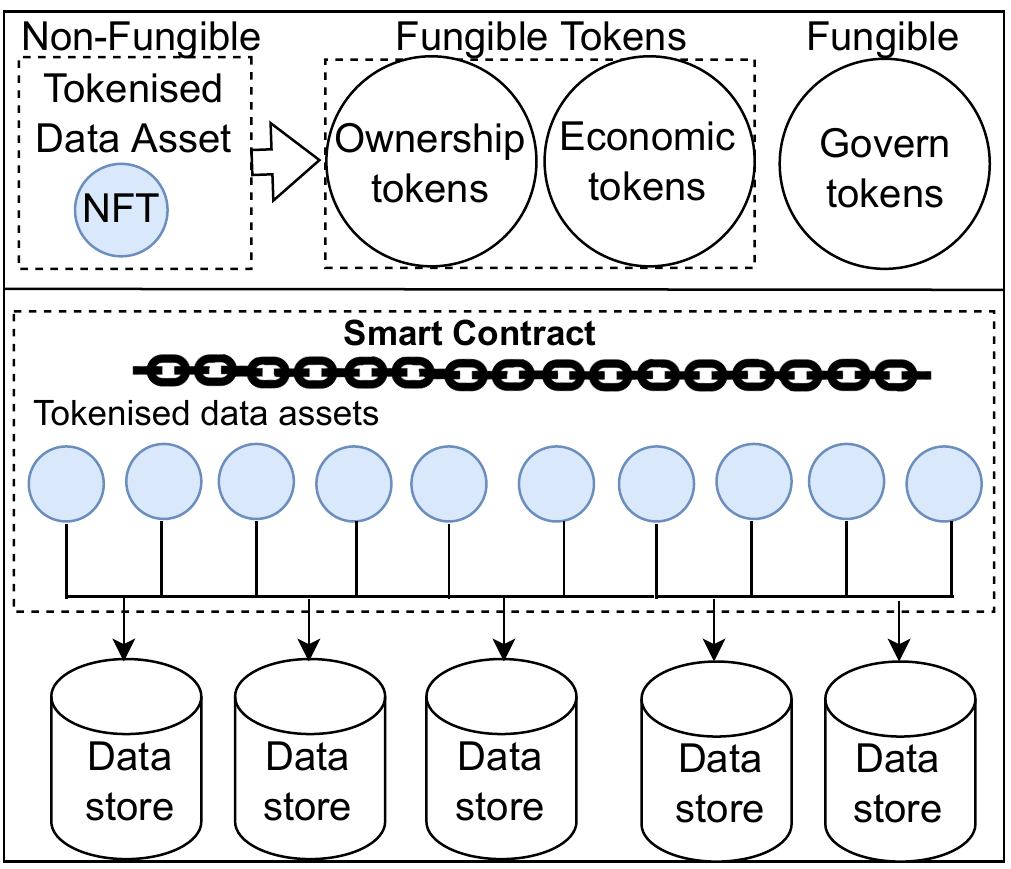}
	\caption{ Each data asset is tokenised into an NFT. An NFT uniquely represents the data asset. Corresponding to each NFT, there are 2 types of fungible tokens, namely, ownership and economic tokens. Governance tokens are used for voting in a DAO. Every NFT points to its storage location in a datastore.}
	\label{fig:shares3}
\end{figure}

\subsection{Ecosystem Tokens}
In our building ecosystem, tokens are used to bridge the gap between the buyer (information consumer) and seller (building owner). It allows scarce but valuable building data made available (via mutual cooperation) by incentivising the owners. 
Our building ecosystem uses 4 types of tokens --- one is non-fungible and the remaining three are fungible (see Fig.~\ref{fig:shares3}).
\begin{compactenum} 
 \item \textit{Non-fungible token.} A NFT uniquely identifies the data asset. The two fungible tokens corresponding to the NFT are ownership and economic tokens. 
 
\item \textit{Ownership tokens.} The sale of ownership (or part of it) may arise when a building or its units change owners and/or when a new building data asset company buys information rights to the building.  Ownership tokens (fungible) are used to facilitate fractional ownership sales.
Any data asset may have one or more owners, and the rule of agreement by a combined majority  of ownership token holders determines the rights (dispute resolution) for the data asset.
Ownership tokens are sold only to other building owners and those who are closely associated with the building (eg., building data asset company).

\item \textit{Economic tokens.} They are fungible and may be sold by building owners to individual/institutional investors on a public blockchain to raise capital or to cover the costs of tokenising the asset and maintaining it (including paying the contractors and for storage). The economic token holders receive royalties for holding this token, and may also be traded on a decentralised exchange for profit.

\item \textit{Govern tokens.} These fungible tokens are unrelated to the NFT and individual asset governance. It will only  be used in overall ecosystem governance via DAO voting (see  section~\ref{sec:dao}).
It is held only by ownership token holders. They are used to ensure the continuity and wellbeing of the ecosystem via majority agreement.

\end{compactenum}

\subsection{Ecosystem bootstrapping and interconnections}
The sequence of steps required to bootstrap the building ecosystem and operate it are shown in Fig.~\ref{fig:ecosystem}. First, the building owners hire contractors to collect building data (Step 1, Fig.~\ref{fig:ecosystem}), who then pass it to owners in the required digital file format (Step 2, Fig.~\ref{fig:ecosystem}). The data is certified and digitally signed by an authorised expert on behalf of the owners (not shown in Fig.~\ref{fig:ecosystem}).  Further, the building owners are not expected to be tech-savvy and use the services of a tokeniser  company/entity  to issue their tokens.
The owner signs a legally binding  contract (Step 3, Fig.~\ref{fig:ecosystem}) with the tokeniser. Based on this contract agreement, the tokeniser mints\footnote{Any tokeniser entity may be used. However, the ecosystem only accepts onboarding of data assets that follow the same set of rules, agreed  by its stakeholders. The contract with the tokeniser is legally binding to prevent them from arbitrarily minting tokens.} (issues) the tokens agreed in the signed contract, and sends it to the owners (Step 4, Fig.~\ref{fig:ecosystem}). The ownership tokens\footnote{Further restrictions may apply to the sale of ownership tokens, as they are only intended to be sold to other building owners, or those closely associated with the physical building.
	This may be a condition of the legal contract between the building owner and tokeniser. Since the buildings are located in a physical jurisdiction, so are  the laws that govern it.
	The DEX only acts as a platform for trustless exchange of tokens.} and economic tokens may be sold on a decentralised exchange (DEX) (Step 5 \& 6, Fig.~\ref{fig:ecosystem}), attached to the building ecosystem.
Further, the information on the datastore pointed to by the data asset (NFT) needs to be searchable without revealing bulk of its information. This is achieved through an online custom search platform that is part of the ecosystem. It allows any interested consumer to query (Step 7a, Fig.~\ref{fig:ecosystem}), for which it returns tokenised data asset (NFT) identifier (Step 7b, Fig.~\ref{fig:ecosystem}).
After the asset identifier is retrieved, the consumer is able to pay the correct NFT contract on the blockchain (Step 7c, Fig.~\ref{fig:ecosystem}). Once payment is confirmed
and the user is authenticated,  access is granted to retrieve the required information from the datastore (Step 7d-7g, Fig.~\ref{fig:ecosystem}).
For the funds paid by the consumer to the tokenised data asset (NFT), 50\%  is  split proportionally between economic token holders and the remaining 50\% is split proportionally between ownership token holders (Step 8a-8b, Fig.~\ref{fig:ecosystem}). Hence, owners are incentivised to hold on to their ownership tokens and work to bring in more monies.
Also, the economic token holders receive a fair share of profits from consumers.
 The  distribution of funds are enforced by the rules in NFT smart contract.

\section{Decentralised Autonomous Organisation}
\label{sec:dao}
\begin{figure}
	\centering
	\includegraphics[width=0.8\linewidth]{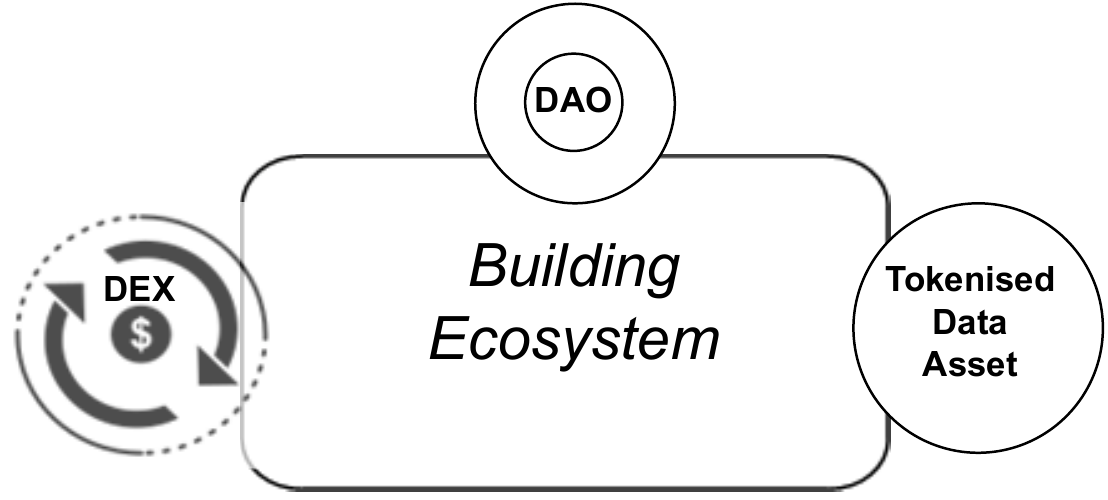}
	\caption{ A DAO is an additional dimension to the building ecosystem to provide governance for the organisation. The DAO is involved in decision making and taking action, when a majority agreement is arrived at.}
	\label{fig:dao}
\end{figure}

For the building ecosystem shown in Fig.~\ref{fig:ecosystem}, we add a new component, namely, the DAO (see Fig.~\ref{fig:dao}) to provide decentralised governance required to keep the system and services running.
Proposals added to the DAO are voted on using the governance tokens held by its stakeholders (see Fig.~\ref{fig:shares3}). A majority vote is used to arrive at an agreement.
Only data asset owners hold DAO governance tokens. The DAO governance tokens are not sold but given to data asset owners by the tokeniser. Further, asset owners are not able to sell DAO governance tokens. When data asset ownership is sold, a corresponding percentage of DAO governance tokens are also transferred to the new asset owner (enforced by the smart contract). It is intended to reduce  speculation and gamification of this token.

\subsection{Caretaker and Strategic DAOs}
\label{ssec:caretakerstrategic}
We define two types of DAOs --- caretaker and strategic.
A caretaker DAO has no strategic or ulterior motive. Its function is to ensure that all the components in the system and its interconnections are running at all times. Suitable action is taken from time to time to manage its upkeep. In general, the proposals in a caretaker DAO are not contentious. It will have an on-chain mechanism to submit a proposal, vote on the proposal, trigger vote tally and winner announcement.
Typically, a treasury system is not part of a caretaker DAO.

\begin{figure}
	\centering
	\includegraphics[width=0.8\linewidth]{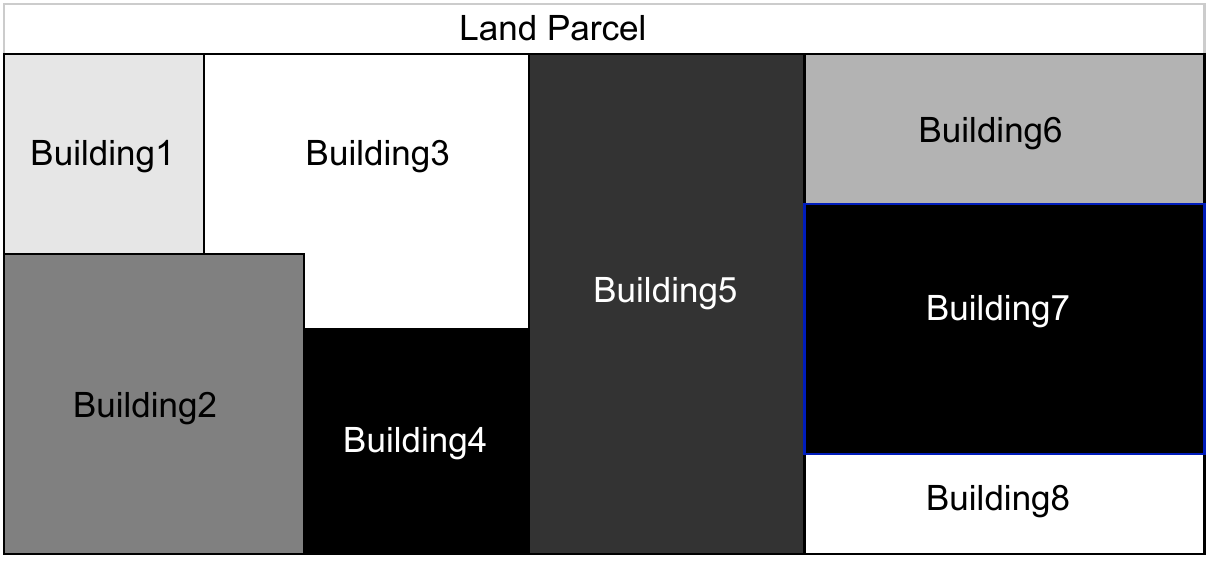}
	\caption{A strategic decision may be made by the DAO to onboard data assets in Building 3 and Building 8, to have a complete information view for the given land parcel. The other buildings  are already part of the ecosystem.  }
	\label{fig:landparcel}
\end{figure}

A strategic DAO is defined  as a value added DAO built to realise strategic goals and includes a treasury.
Consider the example land parcel shown in Fig.~\ref{fig:landparcel}.
The building ecosystem has on boarded all the data assets except for those in buildings 3 and 8.
A strategic proposal may be submitted to bring in the remaining building data assets. This would allow the ecosystem to provide a complete view of the information set, for the given land parcel.
Funds from the DAO treasury may be used to further entice building 3 \& 8 owners  to join by offering to cover their onboarding costs.
This may be contentious as some existing asset owners may feel they did not receive the same incentives. Public deliberation and negotiations may iron out some of the differences before voting on the DAO proposal. However, disagreements are inherent to a strategic DAO and a majority vote is used to accept and fund the proposal using the treasury.

\subsection{Improving DAO Stability}
\label{ssec:improvingstabilitybydesign}
As seen in a strategic DAO, it may not be possible to  eliminate conflict and exit. However, it is possible to reduce the friction by aligning the individual incentives with that of the DAO.
The incentives associated with a token may be built to be least in conflict with other tokens — to help individual participants best position themselves.
The building ecosystem achieves this  by clearly partitioning  governance rights for individual data assets from the broader ecosystem governance rights (determined by DAO). Further, we limit who owns the DAO governance tokens. Our DAO supports the following features.
\begin{compactenum}
	
	\item Clear boundaries. Individual asset owner(s) retain full rights over the ownership of the asset. The DAO has no say in it. This is realised through asset ownership tokens.
	
	\item Clear role. The DAO manages the high-level ecosystem governance only.
	
	\item Alignment of stakeholders in governance. The DAO is governed by data asset owners alone (who are closely associated with the building).
	This prevents anyone from making arbitrary and meaningless proposals and voting on it.
	
	\item Meritocracy. The DAO governance tokens are distributed to data asset owners using meritocratic suffrage  based on positive contribution.
	
	\item Decoupling economics from  governance. The economics (where investors provide capital) are decoupled from  DAO governance via the issuance of economic tokens. Investors  have access to economic tokens alone and do not hold any DAO voting or ownership tokens (unless they are building owners themselves).

	\item Type of DAO. Our building ecosystem uses a default caretaker DAO. If and when a strategic DAO is agreed upon, a small percentage of the monies paid to the data asset will be used to fund the DAO treasury.
\end{compactenum}

\subsection{Hirschman's Tension Triangle \& DAO Tension Quadrilateral}
\label{ssec:tensionquad}

\begin{figure*}
	\centering
	\includegraphics[width=0.9\linewidth]{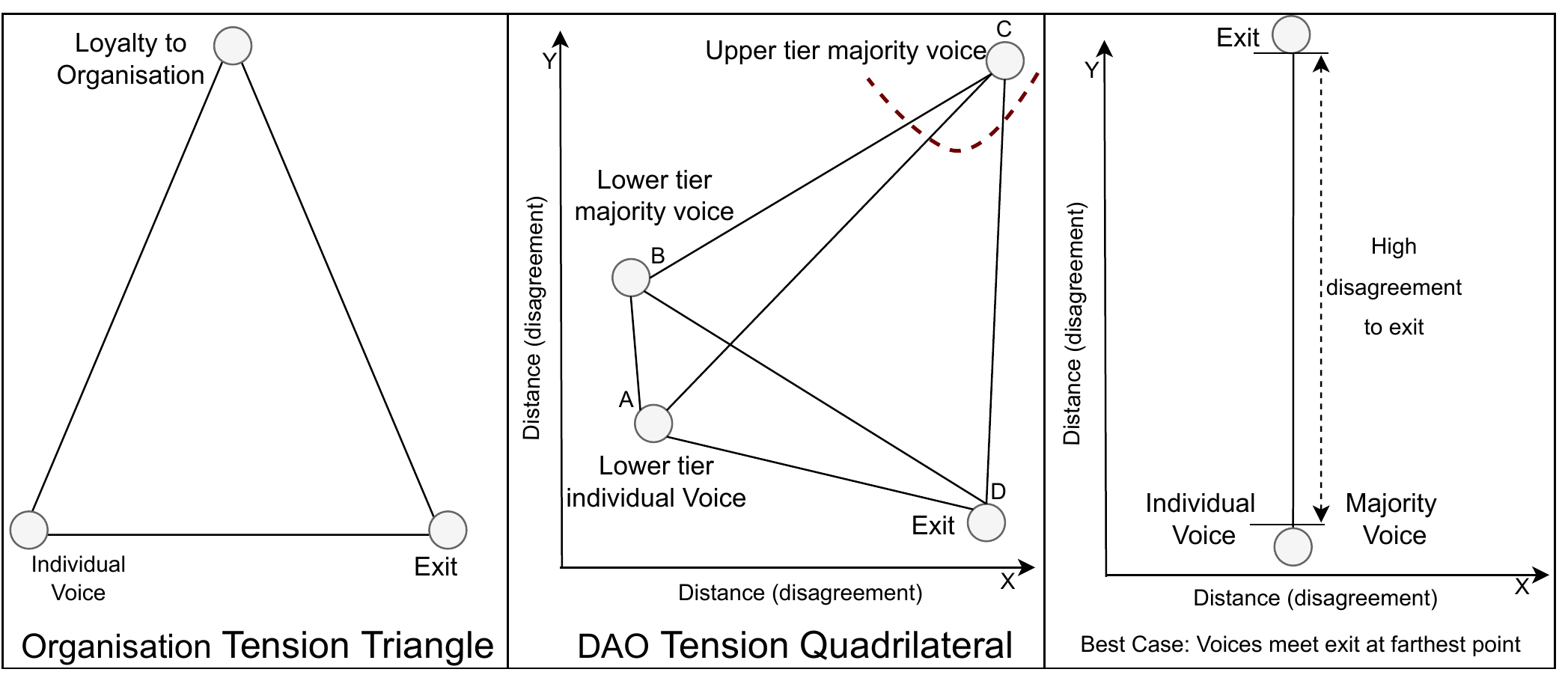}
	\caption{(a). A tension triangle portrays an unstable and mis-aligned set of views between an individual and her organisation, with an option to exit. (b). When a DAO consists of a group of hierarchies, each with voting rights — a tension quadrilateral (proposed) shows the tussle between different tiers, where the higher ranked tier agreement is authorised to block a lower tier agreement. When a DAO has a flat hierarchical structure, the quadrilateral collapses into a triangle with individual voice, majority voice and exit at its 3 vertices. (c). The best case scenario is when individual voice and majority voice are one and the distance to their exit is maximised. }
	\label{fig:tensiontriangle}
\end{figure*}

The organisational tension triangle introduced by Hirschman~\cite{Hirschman} and its effects on various organisational structures including DAOs~\cite{Samman2020} has been explored. It highlights the following struggle to position oneself (an individual) within an organisational tension triangle  (see Fig~\ref{fig:tensiontriangle} (a).) — should an individual be loyal to the organisation and accept the policies imposed by higher management despite her voice not being heard or exit the organisation?
The sides of the triangle are not static and move around  based on the satisfaction of the individual with the organisation.
Unlike organisations where individuals are only able to choose between exit and loyalty, a DAO is uniquely positioned due to the nature of voting rights given to its individual stakeholders.

We draw from Hirschman's work to propose a DAO tension quadrilateral.
Depending on the structure of the DAO (hierarchical or flat),  one of the two cases may appear.

\begin{compactenum}
	
	\item [S1.] Hierarchical structure.  A DAO proposal may be voted on by stakeholders within a lower tier and passed (see Fig~\ref{fig:tensiontriangle} (b).) but requires approval from a higher authoritative tier (agreement via voting). Close proximity of the vertices indicates agreement and distance is synonymous to disagreement. The individual (in Fig~\ref{fig:tensiontriangle} (b).) is a member of the lower tier and its distances to other vertices represents disagreement with the majority voices of both tiers, and how close the individual is to  exiting the DAO.
	
	\item [S2.] Flat structure.  A DAO proposal may be voted by stakeholders within a group and a decision is made based on the outcome of voting. Approval from an upper tier is not required (or non-existent). In this case, the tension quadrilateral in Fig.~\ref{fig:tensiontriangle} (b). collapses into a triangle with individual voice, majority voice and exit at its 3 vertices.
	
\end{compactenum}

In $S1$, if the agreement of  a lower tier is frequently disapproved by a higher tier, then a majority of lower tier stakeholders may collectively exit. In $S2$, an individual may exit if her agreeability to exit is far higher than the agreeability of aligning with the policy of majority stakeholders. The best case is when the goals of the individual and the majority of stakeholders are  aligned to meet at a point, such that their exit distance is high (see Fig~\ref{fig:tensiontriangle} (c.)). The worst case is when the  DAO and exit meet at the same point. In practice, they lie on the vertices of a quadrilateral (or triangle) depending on the organisational structure, pulling or pushing at each other over time.

\subsection{Measuring DAO stability in the Tension Quadrilateral}
\label{ssec:measuring}

\begin{algorithm}[t]
	\label{alg:tensionquad}
	\DontPrintSemicolon
	
	\footnotesize
	\SetKwProg{Fn}{Function}{:}{\KwRet}
	\SetKwFunction{VoteInInterval}{VoteInInterval}
	\SetKwFunction{TallyInInterval}{TallyInInterval}
	\SetKwFunction{UpdateInterval}{UpdateInterval}
	
	Let stakeholders $S=\{s_1,\ldots ,s_j\}$.\\
	Let $\pi_j$ be proof of identity for $j^{th}$ stakeholder.\\
	Let $\pi_{nxtint}$ be proof the trigger to update voting interval is valid.\\
	Let $P_x$ be the $x^{th}$ DAO proposal.\\
	Let choices $C=\{c_1,\ldots ,c_{k=5}\}$.\\
	Let $v_{i,j}$ be the vote choice of $s_j$ in voting interval $i$ for $P_x$.\\
	Let $T=\{t_1,\ldots ,t_{k=5}\}$ be the tally.
	\\Initialise $T \leftarrow 0$,$interval\leftarrow 0$.\tcp{Voting interval to 0}
	
	Input. $P_x, S, C, T, interval,\pi_j$,$\pi_{nxtint}$. \\
	Output. Stability measure ($t_1,\ldots,t_{k=5}$)$\in T$,$\forall$ ($c_1,\ldots,c_{k=5}$)$\in P_x$ in each  voting interval $i$.
	
	\SetKwProg{Fn}{Def}{:}{}
	\Fn{\VoteInInterval{$x, interval, v_{ij}, \pi_j$}}{    
		$i \leftarrow interval$\\
		\If{($isIntervalCurrent(i)$ AND $ verifyVoter(j,\pi_j)$)}
		{
			$vote[x][i][j]\leftarrow  v_{ij}$\tcp{Add to storage}
		}
	}

	\Fn{\TallyInInterval{$x, interval,vote$}}{
		$T\gets 0$\\    
		\If{($isIntervalExisting(interval)==false$)}
		{return -1 \tcp{return failed check}}
		\For{($choice\gets$ 1 to $k$) }
		{
			\For{($stakeholder\gets$ 1 to $j$) }
			{
				\If{$vote[x][interval][stakeholder]==choice$}    
				{
					\tcp{Compute tally for each choice}
					$T[x][interval][choice] = T[x][interval][choice]$+1\\
					
				}
			}
		}
		return $T$
		
	}
	
	\Fn{\UpdateInterval{$nxtint,\pi_{nxtint}$}}{    
		\If{($IntervalUpdateVerify(nxtint,\pi_{nxtint})$)}
		{
			$interval\leftarrow  interval+1$\tcp{Add to storage}
		}
	}

	\caption{Tension  Measure in a DAO Quadrilateral}
\end{algorithm}

Measuring DAO stability helps monitor the health of the organisation. It acts as an early warning system for disagreement and group exit in a DAO leading to its destabilisation/collapse.
Stability may depend on how each individual participant feels about the DAO, and their interactions and relationship between the various tiers in the organisation, at any given point in time.
Stability are also subject to the incentives received via the tokens held, or reactions to external factors such as the narrative~\cite{Riley2015}, inflation~\cite{Shaun2009}, war~\cite{Verdickt2019} and extreme weather~\cite{Kruttli2021}.
To measure a quantity that varies over time, we borrow the idea used in Always-On-Voting (AoV): A  framework for repetitive voting on the blockchain~\cite{Venugopalan2021AoV}. In AoV, an individual is permitted to vote repeatedly, while the effect of their vote is manifested at the end of each voting interval, where votes are tallied and the results published. The intervals repeat over a  bounded but  unpredictable time in the future.
Any DAO stakeholder may privately change their vote at any point in time to show their agreement/disagreement.
The votes are tallied at the end of each interval. The end time of the intervals are random and not known in advance.
The unpredictable timing is incorporated to reduce  the peak-end-effects because individuals may be enticed to vote for an outcome in favour of agenda setters.
A number of blockchain voting algorithms permit  anonymous~\cite{Bulens2011,LaiHOTICN2018}  and/or confidential voting~\cite{Baudron2001,Benabdallah2022}. Any suitable voting algorithm may be plugged-in to the repeated voting framework to measure DAO stability.

A 1-out-of-$k$ voting in a DAO is defined by the 4 tuple ($ S, P, C, T $). The stakeholders ($S$) may vote on any proposal ($P$) for any 1 of their $k$ voting choices ($c_1,c_2,\ldots,c_k\in C$).
A tally ($T$) computes the sum of votes received for each choice  in $C$.
For the DAO quadrilateral (see Fig~\ref{fig:tensiontriangle} (b.)), we may have  a proposal  $P_1$  asking its lower tier stakeholders, ``what is your satisfaction w.r.t to your own tier/group on a scale of 0 to 4? (0  most agreeable and 4 most disagreeable)''. I.e., provide the distance between vertex A and B in Fig~\ref{fig:tensiontriangle} (b.).
For example, a lower tier DAO stakeholder may be asked to indicate her level of agreement with her own tier/group. If her vote choice is 0, her disagreement with her group is 0 (none). Hence, the distance to her group is 0. Both vertices A and B will meet at the same point. This is the best case individual outcome for this proposal. If her vote choice is 4, she is in complete disagreement with her group. Vertices A and B will be at the farthest distance from each other.
Another proposal $P_2$ may ask a lower tier individual, ``what is your satisfaction in dealing with a higher tier/group in your organisation on a scale of 0 to 4? (0  most agreeable and 4 most disagreeable)''. I.e., provide the distance between vertex A and C in Fig~\ref{fig:tensiontriangle} (b.).
A third proposal $P_3$  may ask ``how close are you as a lower tier stakeholder  to exiting the DAO? (0  closest to exit and 4 farthest)''. I.e., provide the distance between vertex A and D in Fig~\ref{fig:tensiontriangle} (b.).
A vote choice of 0 implies the stakeholder is in complete agreement with the proposal of exiting the DAO and a vote choice of 4 implies complete disagreement with exiting the DAO.

For each proposal, we provide 5 voting choices ($C=c_1,c_2,\ldots,c_5$). The choices represent a linear scale arranged in ascending order. I.e., ($c_1 \rightarrow 0,\ldots,c_5 \rightarrow 4$).
The Always-On-Voting  framework is incorporated to quantify and measure DAO stability in the proposed tension quadrilateral (see Alg. \ref{alg:tensionquad}). It involves 3 main functions.
The function $VoteInInterval$ allows an individual in the current voting interval to vote  on proposal $P_x$, for her desired choice. An individual is also permitted to update her vote. However, when the current voting interval ends, and the interval is updated, votes in previous intervals can no longer be changed.  The function $TallyInInterval$ is called to tally the votes in any given interval. It returns a tally as a measure of stability, for the given proposal in the DAO (corresponding to the given voting interval). The  function $Updateinterval$ is used to update the election to its next voting interval.
A proof $\pi_j$ is used to prove voting eligibility  of the $j^{th}$ stakeholder. Another proof $\pi_{nxtint}$ is used to trigger the next voting interval. Alg. \ref{alg:tensionquad} verifies these proofs for its correctness before taking the appropriate action.

The voting in Alg. \ref{alg:tensionquad} supports one vote per individual. This is useful to find the tension across a 
number of individual voters in the DAO. In addition, a weighted voting system may be considered to measure power holder agreement (those who wield significant power in the DAO), wherein some voters have more voting power than others (based on the number of governance tokens held by an individual).  
Its incorporation and technical details are left as future work.

\section{Discussion}
\label{sec:discussion}

The incentives gave stakeholders a reason to participate in the ecosystem and tokens acted as a carrier for coordinating its activities.
The security provided by the blockchain ensured the activities were not disrupted. The DAO took action via a majority consensus, when changes were required.
Blockchain tokens and  a DAO were used to  meet some of the data asset requirements (R1-R5) in Section~\ref{sec:prelims}. The non-fungibility property of a NFT was used to prevent arbitrary token duplication (R1) but data store security~\cite{Dipanjan2022} remains its weakest link. The security provided by a blockchain was used to give exclusive control  (R2) to ownership token holders.
Data store held building information was certified by accredited experts, to partly comply with regulatory requirements (R3).
All blockchain log entries are immutable and digitally signed by its creator. This aids  to meet the audit requirements (R4). The requirement R5 was to provide services by ensuring system functionality.
For this, a DAO was used to  manage ecosystem functionality and changes.

\section{Conclusions}
\label{sec:conclusions}
We presented our use case on building data assets to show that tokenomics and governance go hand-in-hand to reduce the tension between stakeholders  participating in the building ecosystem.
We identified group exit and DAO destabilisation as existential threats to our building ecosystem.
Unlike in a caretaker DAO, tension is inherent in a strategic DAO. Though it cannot be entirely eliminated, measures were taken to reduce it.
This was achieved by reducing the misalignment between the tokens, by clearly partitioning its roles, responsibilities, incentives and ownership.
Further, we modified the Hirschman's tension triangle to adapt it to a DAO and created the DAO tension quadrilateral.
Additionally, we proposed a tool to measure and quantify tension between stakeholders and tiers/groups in the DAO quadrilateral.
This measure may be used to find common ground, and acts as an early warning system for DAO destabilisation and exit.

\section*{Acknowledgment}

This research is supported by the National Research Foundation, under its Campus for Research Excellence and Technological Enterprise (CREATE) Programme.

\bibliographystyle{IEEEtran}
\bibliography{ref-reduced}

\begin{thebibliography}{10}
\providecommand{\url}[1]{#1}
\csname url@samestyle\endcsname
\providecommand{\newblock}{\relax}
\providecommand{\bibinfo}[2]{#2}
\providecommand{\BIBentrySTDinterwordspacing}{\spaceskip=0pt\relax}
\providecommand{\BIBentryALTinterwordstretchfactor}{4}
\providecommand{\BIBentryALTinterwordspacing}{\spaceskip=\fontdimen2\font plus
\BIBentryALTinterwordstretchfactor\fontdimen3\font minus
  \fontdimen4\font\relax}
\providecommand{\BIBforeignlanguage}[2]{{%
\expandafter\ifx\csname l@#1\endcsname\relax
\typeout{** WARNING: IEEEtran.bst: No hyphenation pattern has been}%
\typeout{** loaded for the language `#1'. Using the pattern for}%
\typeout{** the default language instead.}%
\else
\language=\csname l@#1\endcsname
\fi
#2}}
\providecommand{\BIBdecl}{\relax}
\BIBdecl

\bibitem{SuwitoTSDTUS22}
M.~H. Suwito, B.~A. Tama, B.~Santoso, S.~Dutta, H.~Tan, Y.~Ueshige, and
  K.~Sakurai, ``A systematic study of bulletin board and its application,'' in
  \emph{{ASIA} {CCS} '22: {ACM} Asia Conference on Computer and Communications
  Security, Nagasaki, Japan, 30 May 2022 - 3 June 2022}, Y.~Suga, K.~Sakurai,
  X.~Ding, and K.~Sako, Eds.\hskip 1em plus 0.5em minus 0.4em\relax {ACM},
  2022, pp. 1213--1215.

\bibitem{Szabo1997}
\BIBentryALTinterwordspacing
N.~Szabo, ``The idea of smart contracts,'' 1997. [Online]. Available:
  \url{https://nakamotoinstitute.org/the-idea-of-smart-contracts/}
\BIBentrySTDinterwordspacing

\bibitem{governance2022}
\BIBentryALTinterwordspacing
A.~Kiayias and P.~Lazos, ``Sok: Blockchain governance,'' \emph{CoRR}, vol.
  abs/2201.07188, 2022. [Online]. Available:
  \url{https://arxiv.org/abs/2201.07188}
\BIBentrySTDinterwordspacing

\bibitem{Lamberty2020}
\BIBentryALTinterwordspacing
R.~Lamberty, D.~de~Waard, and A.~Poddey, ``Leading digital socio-economy to
  efficiency -- a primer on tokenomics,'' 2020. [Online]. Available:
  \url{https://arxiv.org/abs/2008.02538}
\BIBentrySTDinterwordspacing

\bibitem{Hunhevicz2022}
J.~J. Hunhevicz, ``Exploring the potential of blockchain and cryptoeconomics
  for the construction industry,'' Doctoral Thesis, ETH Zurich, 2022.

\bibitem{Barone2022}
A.~Barone, ``What is an asset? definition, types, and examples,'' 2022,
  \url{https://www.investopedia.com/terms/a/asset.asp }.

\bibitem{Honic2021}
\BIBentryALTinterwordspacing
M.~Honic, I.~Kovacic, P.~Aschenbrenner, and A.~Ragossnig, ``Material passports
  for the end-of-life stage of buildings: Challenges and potentials,''
  \emph{Journal of Cleaner Production}, vol. 319, p. 128702, 2021. [Online].
  Available:
  \url{https://www.sciencedirect.com/science/article/pii/S0959652621029012}
\BIBentrySTDinterwordspacing

\bibitem{Sanchez2021}
\BIBentryALTinterwordspacing
B.~Sanchez, C.~Rausch, C.~Haas, and T.~Hartmann, ``A framework for bim-based
  disassembly models to support reuse of building components,''
  \emph{Resources, Conservation and Recycling}, vol. 175, p. 105825, 2021.
  [Online]. Available:
  \url{https://www.sciencedirect.com/science/article/pii/S0921344921004341}
\BIBentrySTDinterwordspacing

\bibitem{Cetin2021}
\BIBentryALTinterwordspacing
S.~Çetin, C.~De~Wolf, and N.~Bocken, ``Circular digital built environment: An
  emerging framework,'' \emph{Sustainability}, vol.~13, no.~11, p. 6348, Jun
  2021. [Online]. Available: \url{http://dx.doi.org/10.3390/su13116348}
\BIBentrySTDinterwordspacing

\bibitem{Heiko2022}
H.~Aydt, G.~Habert, D.~Hall, S.~Hellweg, P.~Herthogs, v.~Richthofen, Aurel, and
  R.~Stouffs, \emph{\BIBforeignlanguage{en}{Action 74: Enable Sustainable
  Material Flows}}.\hskip 1em plus 0.5em minus 0.4em\relax Zurich: Lars Müller
  Publishers, 2022, pp. 200 -- 201, 308 -- 311, evidence 74: Enable Sustainable
  Material Flows, p. 308-311.

\bibitem{Msisac2022}
MSISAC, ``Understanding denial-of-service attacks,'' 2022,
  \url{https://www.cisa.gov/uscert/ncas/tips/ST04-015 }.

\bibitem{Georgina2015}
G.~Hurst, ``If a custody bank fails, should clients fear for their assets?''
  2015,
  \url{https://www.institutionalinvestor.com/article/b14z9y5332m1mt/if-a-custody-bank-fails-should-clients-fear-for-their-assets
  }.

\bibitem{Hirschman}
A.~O. Hirschman, \emph{Exit, voice, and loyalty: responses to decline in firms,
  organizations, and states}.\hskip 1em plus 0.5em minus 0.4em\relax Harvard
  University Press, 1970.

\bibitem{Samman2020}
G.~Samman and D.~Freuden, ``{DAO: A Decentralized Governance Layer for the
  Internet of Value},'' Tech. Rep., May 2020.

\bibitem{Riley2015}
T.~Riley and B.~Luippold, ``Managing investors' perception through strategic
  word choices in financial narratives,'' \emph{Journal of Corporate Accounting
  and Finance}, vol.~26, 07 2015.

\bibitem{Shaun2009}
\BIBentryALTinterwordspacing
S.~K. Roache and A.~P. Attie, ``Inflation hedging for long-term investors,''
  \emph{IMF Working Papers}, vol. 2009, no. 090, p. A001, 2009. [Online].
  Available:
  \url{https://www.elibrary.imf.org/view/journals/001/2009/090/article-A001-en.xml}
\BIBentrySTDinterwordspacing

\bibitem{Verdickt2019}
G.~Verdickt, ``The effect of war risk on managerial and investor behavior:
  Evidence from the brussels stock exchange in the pre-1914 era,'' \emph{SSRN
  Electronic Journal}, 10 2019.

\bibitem{Kruttli2021}
M.~S. Kruttli, B.~Roth~Tran, and S.~W. Watugala, ``Pricing poseidon: Extreme
  weather uncertainty and firm return dynamics,'' in \emph{Federal Reserve Bank
  of San Francisco Working Paper Series.}, 2021.

\bibitem{Venugopalan2021AoV}
\BIBentryALTinterwordspacing
S.~Venugopalan, I.~Stančíková, and I.~Homoliak, ``Always on voting: {A}
  framework for repetitive voting on the blockchain,'' \emph{CoRR}, vol.
  abs/2107.10571, 2021. [Online]. Available:
  \url{https://arxiv.org/abs/2107.10571}
\BIBentrySTDinterwordspacing

\bibitem{Bulens2011}
\BIBentryALTinterwordspacing
P.~Bulens, D.~Giry, and O.~Pereira, ``Running {Mixnet-Based} elections with
  helios,'' in \emph{2011 Electronic Voting Technology Workshop/Workshop on
  Trustworthy Elections (EVT/WOTE 11)}.\hskip 1em plus 0.5em minus 0.4em\relax
  San Francisco, CA: USENIX Association, Aug. 2011. [Online]. Available:
  \url{https://www.usenix.org/conference/evtwote-11/running-mixnet-based-elections-helios}
\BIBentrySTDinterwordspacing

\bibitem{LaiHOTICN2018}
W.-J. Lai, Y.-c. Hsieh, C.-W. Hsueh, and J.-L. Wu, ``Date: A decentralized,
  anonymous, and transparent e-voting system,'' in \emph{2018 1st IEEE
  International Conference on Hot Information-Centric Networking (HotICN)},
  2018, pp. 24--29.

\bibitem{Baudron2001}
O.~Baudron \emph{et~al.}, ``Practical multi-candidate election system,'' in
  \emph{PODC '01}.\hskip 1em plus 0.5em minus 0.4em\relax New York, NY, USA:
  ACM, 2001, pp. 274--283.

\bibitem{Benabdallah2022}
A.~Benabdallah, A.~Audras, L.~Coudert, N.~El~Madhoun, and M.~Badra, ``Analysis
  of blockchain solutions for e-voting: A systematic literature review,''
  \emph{IEEE Access}, vol.~10, pp. 70\,746--70\,759, 2022.

\bibitem{Dipanjan2022}
D.~Das, P.~Bose, N.~Ruaro, C.~Kruegel, and G.~Vigna, ``{Understanding} security
  issues in the nft ecosystem,'' in \emph{CCS '22}.\hskip 1em plus 0.5em minus
  0.4em\relax ACM, 2022.

\end{thebibliography}

\end{document}